\documentclass[12pt, a4paper]{article} 
\usepackage{a4wide}
\usepackage{amsmath,amssymb} 
\usepackage{float}

\usepackage[english]{babel}

\usepackage{hyperref}

\newcommand{\e}{\ensuremath{{\mathrm{e}}}}
\newcommand{\ei}{\ensuremath{\mathrm{e}^{-\mu}}}
\newcommand{\eii}{\ensuremath{\mathrm{e}^{-2\mu}}}
\newcommand{\eiii}{\ensuremath{\mathrm{e}^{-3\mu}}}
\newcommand{\eiiii}{\ensuremath{\mathrm{e}^{-4\mu}}}
\newcommand{\ci}{\ensuremath{\cosh \mu}} 
\newcommand{\si}{\ensuremath{\sinh \mu}} 

\newcommand{\siii}{\ensuremath{\sinh^3 \mu}}

\newcommand{\SUM}{\ensuremath{\sum_{\mathbf{k}}{}^{'}}}
\newcommand{\SUMV}{\ensuremath{\frac{1}{V_s}\sum_{\mathbf{k}}{}^{'}}}

\newcommand{\CS}{\ensuremath{\left| \sum_{i=1}^{d-1} \mathrm{e}^{i
        \mathbf{k_i}}(1-\cos \mathbf{k_i})\right|^2}}

\newcommand{\Cisq}{\ensuremath{\sum_{i=1}^{d-1}(1-\cos \mathbf{k_i})^2}}

\newcommand{\Sm}{\mathcal{S}} 
\newcommand{\Em}{\mathcal{E}}
\newcommand{\Cm}{\mathcal{C}}

\allowdisplaybreaks[1] \numberwithin{equation}{section}

\begin{document}

\thispagestyle{empty} \parskip=12pt \raggedbottom

\vspace*{1cm}
\begin{center}
  {\LARGE The rotator spectrum in the delta-regime of the O($n$) effective
    field theory in 3 and 4 dimensions}

  \vspace{1cm}
  Ferenc~Niedermayer and Christoph~Weiermann \\[3mm]
  Albert Einstein Center for Fundamental Physics \\

  Institute for Theoretical Physics, Bern University \\
  Sidlerstrasse 5, CH-3012 Bern, Switzerland \\
  \vspace{0.5cm}
  
  \nopagebreak[4]
  
  \begin{abstract}
    The low lying spectrum of the O($n$) effective field theory is calculated
    in the delta-regime in 3 and 4 space-time dimensions using lattice
    regularization to NNL order. It allows, in particular, to determine, using
    numerical simulations in different spatial volumes, the pion decay
    constant $F$ in QCD with 2 flavours or the spin stiffness $\rho$ for an
    antiferromagnet in $d=2+1$ dimensions.
  \end{abstract}

\end{center}

\eject

\section{Introduction}
The low energy phenomena in systems with spontaneously broken symmetry are
governed by the dynamics of the Goldstone bosons.  This can be described by an
effective field theory, and the calculations could be performed by chiral
perturbation theory (ChPT) \cite{Weinberg,Gasser:1983yg}.  The effective
action contains low energy constants (LEC) determined by the underlying
microscopic theory.  The physical quantities can be systematically expanded in
powers of momenta, or (as in the case of our interest here) inverse box size.
In numerical simulations one can place the system into a space-time box of
size $L_t\times L_s^{d-1}$ and study the dependence of different quantities on
the box size. Comparing the data with theoretical predictions one can
determine the corresponding LEC's.

There are important cases when the order parameter of the spontaneous symmetry
breaking is an O($n$) vector.  These include QCD with two light quarks (4d
O(4) case), and antiferromagnetic layers (3d O(3) case).
 
The different regimes of such systems in a finite box have been systematized
by Leutwyler \cite{Leutwyler:1987ak}.  In particular, for the case of no
explicit symmetry breaking (zero quark mass in QCD) one can distinguish the
$\epsilon$-regime (``cubic geometry''), where $L_t \sim L_s$, and the
$\delta$-regime (``cylindrical geometry'') where $L_t \gg 1/m(L_s) \sim F^2
L_s^3 \gg L_s$.  (Note that the expansion parameter of ChPT in this case is
$1/(F^2 L_s^2)$ and must be small enough.)

In this work we calculate the low lying spectrum in a finite spatial box in
the O($n$) effective theory (i.e. we consider the $\delta$-regime) with no
explicit symmetry breaking, in 3 and 4 space time dimensions,
using lattice regularization.
It has been shown in \cite{Leutwyler:1987ak} that
in the leading order the spectrum is given by the quantum mechanical O($n$)
rotator with moment of inertia $\Theta=(n-1)/2 \cdot F^2 L_s^3$ 
(the ``angular momentum'' being the O($n$) isospin).  
In the case of QCD the constant $F$ is the pion decay constant.  
The next-to-leading (NL) order term of the expansion in $1/(F^2 L_s^2)$ 
has been calculated in \cite{Hasenfratz:1993vf}.  In the calculation
one considered an infinitely long lattice in the time direction, separating
the spatially constant slow modes and the fast modes, and integrating out the
latter.  The resulting effective Lagrangian for the slow modes is then that of
an O($n$) rotator, with a modified moment of inertia.  The NLO correction
turned out to be large: at $L_s=2.5\,\mathrm{fm}$ it is still $\sim 40\%$. 
Therefore it was important to calculate the NNLO term.  This has been
done recently for the 4d, O(4) case by P.~Hasenfratz \cite{Hasenfratz:2009mp}
by a method similar to the one used in \cite{Hasenfratz:1993vf}, except that
in \cite{Hasenfratz:2009mp} dimensional regularization (DR) was used.  The
NNLO term contains two new LEC's, $\Lambda_1$ and $\Lambda_2$, and is
estimated to be $-5\%$ at $L_s=2.5\,\mathrm{fm}$.

Since the calculation using DR with the time-dependent slow modes was quite
involved, we have chosen to calculate the same quantity by a different method,
following the calculation of the small-volume mass gap in 2d O($n$) non-linear
sigma-model by L\"uscher, Weisz, Wolff \cite{Luscher:1991wu}, using lattice
regularization, and we considered both the 3d and 4d cases for general O($n$).

As mentioned above, the effective theory has also been successfully applied in
condensed matter physics.  In particular one can perform numerical simulations
in the microscopic theory of the spin $\frac12$ antiferromagnetic Heisenberg
model and measure different quantities (like staggered susceptibility, etc.)
with an impressive precision \cite{Wie94,Bea96}.  Comparing these with the
results of the effective field theory one can obtain the LEC's 
(spin stiffness, etc) to high accuracy of order $10^{-3}$ 
(for a recent paper see \cite{Ger09}).  
Given the high accuracy, in this case even a small NNLO term
could have an important effect.

The rest of the paper is organized as follows.  Section \ref{sct:Action}
discusses the different terms in the effective action needed to this order,
section \ref{sct:ct} recapitulates the method of \cite{Luscher:1991wu} to
obtain the perturbative expansion for the mass gap. In section 4 we describe
the numerous checks of our calculations, and section 5 gives the general
expression for the mass gap.  The numerical values and the renormalization of
the couplings for $d=4$ and $d=3$ dimensions are given in sections 6 and 7,
respectively.  Section 8 contains our conclusions.  Some further details of
the calculations are delegated to the Appendices.

\section{The effective action} \label{sct:Action} 
We consider the following
effective action in $3$ and $4$ space-time dimensions
\begin{equation} \label{action1} A= A_2+A_4+\ldots\,,
\end{equation}
where the leading term is
\begin{equation} \label{A2}
  A_2=\frac{1}{2\lambda_0^2}\sum_{x} \partial_\mu\mathbf{S}_x\cdot
  \partial_\mu\mathbf{S}_x,
  \quad x=(x_0,\mathbf{x}),  \quad \mathbf{S}_x^2=1.
\end{equation}
Here $\partial_\mu$ denotes the standard forward lattice derivative, and the
$A_4$ part containing 4-derivative terms will be specified later.  Note that
we do not consider here an explicit O($n$) symmetry breaking term.

The non-linear sigma model given by $A_2$ is non-renormalizable in $d>2$
dimensions hence one has to consider $A$ as an effective low energy action,
with corresponding (infinitely many) low energy constants (LEC).  The mass gap
(and the energy of lowest states) in a box $L_s^{d-1}$ can be expanded in
inverse powers of the box size $L_s$, and to a given order in $1/L_s$ only
finite number of LEC's appear. For our case of NNLO corrections we need only
terms up to four derivatives in $d=4$.

The action \eqref{action1} in $d=3$ describes e.g. the low energy behaviour of
the antiferromagnetic spin $\frac12$ quantum Heisenberg model, while in $d=4$,
and $n=4$ it is the effective action of QCD with two massless quarks.

The dimensionless lattice coupling $\lambda_0$ in eq.~\eqref{A2} is expressed
through the dimensionful lattice bare parameters of the corresponding theories
as
\begin{align}
  \lambda_0^2 &=\frac{1}{\rho_0 a} \,, \qquad \;\;\, \text{for } d=3\,, \\
  \lambda_0^2 & =\frac{1}{F_0^2 a^2} \,, \qquad \text{for } d=4\,,
\end{align}
where $\rho_0$ and $F_0$ denote the bare spin stiffness and bare pion decay
constant, respectively, and $a$ is the lattice spacing.\footnote{In most
  expressions we use the convention ``$a=1$'', but restore the $a$ factors
  when it is useful for understanding.}  We parametrize the spin vector with
the ``pion'' fields
\begin{equation}
  \mathbf{S}=\left(\lambda_0\vec{\pi}, \sqrt{1-\lambda_0^2\vec{\pi}^2}\right),
  \quad \vec{\pi}=\left(\pi^1,\pi^2,\ldots,\pi^{n-1}\right).
\end{equation}
The pion fields reflect the perturbative fluctuations around the magnetization
axis. We expand in the $\vec{\pi}$-fields keeping the volume 
$V_d=L_t\times L_s^{d-1}$ finite assuming a cylindrical geometry, 
$L_t \gg L_s$.  As long as the volume is finite there is a slowly moving 
global mode which corresponds to the direction of the magnetization.  
This mode is treated nonperturbatively in
the the path integral \cite{Hasenfratz:1984jk}.

The partition function takes the form
\begin{equation} \label{eq:Z} Z = {\mathcal{N}}\prod_n \int d\vec{\pi}_n
  \,\delta\left(\frac{1}{V_d}\sum_x \vec{\pi}_x\right) \,
  \mathrm{e}^{-A_\mathrm{eff}[\vec{\pi}]}
\end{equation}
where
\begin{equation}
  A_\mathrm{eff}[\vec{\pi}]=A[\vec{\pi}]
  +A_{\mathrm{measure}}[\vec{\pi}]+A_{\mathrm{zero}}[\vec{\pi}].
\end{equation}
The first term is the action of \eqref{action1} (expressed in terms of
$\vec{\pi}$-fields). The second term is caused by the measure from the change
to $\vec{\pi}$-variables
\begin{equation}
  A_{\mathrm{measure}}[\vec{\pi}]=
  \sum_x \ln\left( 1-\lambda_0^2\vec{\pi}^2_x\right)^\frac{1}{2}.
\end{equation}
The irrelevant factor ${\mathcal{N}}$ and the last term together with the
delta function arise when we separate the global zero mode and integrate it
out in the path integral,
\begin{equation}
  A_{\mathrm{zero}}[\vec{\pi}]=
  -(n-1)\ln \sum_x \left( 1-\lambda_0^2\vec{\pi}^2_x\right)^\frac{1}{2}.
\end{equation}
As a consequence of the delta function in \eqref{eq:Z} we must leave out the
$p=(0,\mathbf{0})$ zero mode in the momentum decomposition of $\vec{\pi}$.

For the $d=4$ case at NNLO we also need the $A_4$ term in eq.~\eqref{action1}.
The most general 4-derivative interaction consistent with the lattice
symmetries is given by (see e.g. \cite{Balog:2009np})
\begin{equation}
  A_4=\sum_{i=2}^5 \frac{g_4^{(i)}}{4} \left(A_4^{(i)}
    -c^{(i)} \sum_x \partial_\mu \mathbf{S}_x \cdot
    \partial_\mu \mathbf{S}_x \right) ,\label{eq:A4total}
\end{equation}
where
\begin{align}
  &A_4^{(1)}=\sum_x \partial_\mu \partial_\mu^*\mathbf{S}_x \cdot
  \partial_\nu \partial_\nu^* \mathbf{S}_x, \\
  &A_4^{(2)}=\sum_x (\partial_\mu\mathbf{S}_x \cdot
  \partial_\mu \mathbf{S}_x)^2 ,\\
  &A_4^{(3)}=\sum_x (\partial_\mu\mathbf{S}_x \cdot
  \partial_\nu \mathbf{S}_x)^2, \\
  &A_4^{(4)}=\sum_x \sum_\mu(\partial_\mu\mathbf{S}_x \cdot
  \partial_\mu
  \mathbf{S}_x)^2-\frac{1}{d+2}\left(A_4^{(2)}+2A_4^{(3)}\right)\,,
  \label{eq:A44}\\
  &A_4^{(5)}=A_4^{(5a)}-\frac{1}{d+2}\left(2A_4^{(5b)}+A_4^{(5c)}
  \right), \label{eq:A45}
\end{align}
and
\begin{align}
  &A_4^{(5a)}=\sum_x \sum_\mu \partial_\mu \partial_\mu^* \mathbf{S}_x \cdot
  \partial_\mu \partial_\mu^*\mathbf{S}_x, \\
  &A_4^{(5b)}=A_4^{(1)},\\
  &A_4^{(5c)}=\sum_x \partial_\mu \partial_\mu \mathbf{S}_x
  \cdot \partial_\nu \partial_\nu\mathbf{S}_x .
\end{align}
We use the standard forward ($\partial$) and backward ($\partial^*$) lattice
derivatives. Some comments are in order here:
\begin{itemize}
\item The interaction $A_4^{(1)}$ is redundant at the order of our
  calculation. Transforming variables in the path integral as 
  $\mathbf{S}\to (\mathbf{S}+\alpha \Box \mathbf{S}) 
  /|\mathbf{S}+\alpha \Box \mathbf{S}|$
  the leading action exhibits a contribution proportional to
  $A_4^{(1)}$. Choosing the parameter $\alpha$ we can absorb the contributions
  of this interaction \cite{Hasenfratz:1989pk}.

\item In ChPT of QCD with $N_f=2$ flavours (in Minkowski space) the standard
  notation for the low energy constants is $l_i$ \cite{Gasser:1986vb}.
  Following \cite{Hasenfratz:1989pk} for the O($n$) case (in Euclidean space)
  we use the convention $g_4^{(i)}/4$ for the couplings.  They are related by
  \begin{equation} \label{g23-l12} 
    g_4^{(2)}=-4 l_1\,,\quad g_4^{(3)}=-4 l_2
    \,.
  \end{equation}
  The minus sign comes from going from Minkowski to Euclidean space.  Note
  also that the convention for numbering the operators is different.

\item The operators 4 and 5 are absent in dimensional regularization.  They
  are needed to restore Lorentz symmetry on the lattice and their coefficients
  are fixed by this requirement.  The subtracted pieces stem from the fact
  that we construct first the symmetric, traceless 4-index tensors, and take
  the sum $\sum_\mu t_{\mu\mu\mu\mu}$ \cite{Balog:2009np}.

\item According to equation \eqref{eq:A4total} we subtract a term proportional
  to the leading action $A_2$ from each of the 4-derivative interactions.  The
  coefficients $c^{(i)}$ serve to remove the power-like divergence $1/a^4$
  from the contribution of the corresponding operator.  The subtracted
  operators renormalize then multiplicatively.  This is discussed later.
\end{itemize}

\section{Correlation function}\label{sct:ct}
We extract the mass gap from the correlation function
\begin{equation}
  C(t)=\frac{1}{V_s^2} \sum_{\mathbf{x},\mathbf{y}} 
  \langle \mathbf{S}_{x_0,\mathbf{x}}\cdot \mathbf{S}_{y_0, \mathbf{y}} \rangle,
  \quad t=|y_0-x_0|\,, \label{corr1}
\end{equation}
where $V_s=L_s^{d-1}$ is the spatial volume.  We follow closely the method of
ref.\cite{Luscher:1991wu} developed to obtain the small-volume mass gap
$m(L_s)$ for the 2d O($n$) nonlinear sigma-model.

We apply perturbation theory at fixed finite cylindrical volume 
$L_t\times L_s^{d-1}$ and following \cite{Luscher:1991wu} we impose 
free boundary conditions in the time directions\footnote{with periodic 
  b.c. in the spatial directions}.  
This guarantees that we project onto states with zero isospin
(and zero total momentum) at the two boundaries.  Except for the ground state
all these states (``scattering states'') have energies $\gtrsim 4\pi/L_s$.
The correlation function drops off as
\begin{equation} \label{Ctm} 
  C(t)= A\mathrm{e}^{-m(L_s) t}
  +{\mathcal{O}}\left(\mathrm{e}^{-\frac{4\pi}{L_s}t}\right)\,,
\end{equation}
where $m(L_s)=E_1-E_0$ is the mass gap, the difference between the lowest
energies in the $l=1$ and $l=0$ isospin sectors.  For small $\lambda_0$ the
mass gap is $m(L_s) \propto \lambda_0^2/ V_s$ hence for 
$\lambda_0^2 \ll V_s/L_t$ the system is nearly completely polarized --
 the spins fluctuate only slightly around the direction of the total 
magnetization.  Therefore for fixed $L_s,L_t$ one can use perturbation 
theory in $\lambda_0$ to calculate the correlation 
function\footnote{The notations are analogous to those of
  ref.~\cite{Luscher:1991wu}, except that there a symmetric set-up with
  $x_0=-\tau$, $y_0=\tau$, $t=2\tau$ is used.}
\begin{equation} \label{Ct} 
  C(t)=1+\sum_{i=0}^{\infty}
  \left(\lambda_0^2\right)^{i+1} C_i\left( \frac{t}{L_s} \right) \,.
\end{equation}
Inserting the expansion of the mass gap $m(L_s)$ into eq.~\eqref{Ctm} one
concludes that the coefficients $C_i(t/L_s)$ are (up to exponentially
decreasing terms) polynomials of order $i+1$ in the time $t$
\begin{equation} \label{Cit} 
  C_i\left(\frac{t}{L_s}\right)= \sum_{k=0}^{i+1}
  C_{ik} \cdot \left(\frac{t}{L_s}\right)^k
  +{\mathcal{O}}\left(\mathrm{e}^{-\frac{4\pi}{L_s}t}\right) \,.
\end{equation}

As seen from eq.~\eqref{Ctm} (neglecting the exponentially small terms in $t$)
the expansion of $\log C(t)$ is linear in $t$, the higher order terms in $t$
should cancel.  To obtain the mass gap to NNLO only the coefficients $C_{00}$,
$C_{01}$, $C_{10}$, $C_{11}$, and $C_{21}$ are needed, while $C_{12}$,
$C_{22}$, and $C_{23}$ are useful to check the absence of $t^2$ and $t^3$
terms in $\log C(t)$.

Observe that the use of free b.c. in the time direction is essential here.
Using periodic b.c. also in this direction, the transitions $l\to l+1$ 
between states of higher isospins with energies 
$E_l \approx l(l+n-2) \lambda_0^2 / (2 V_s)$ would also contribute to 
the perturbative expansion of $C(t)$, making the method unpractical 
for the determination of the mass gap.

Note that the method of ref.~\cite{Luscher:1991wu} is essentially an
$\epsilon$-regime expansion in a very elongated cylindrical volume, since the
spins are strongly correlated over the whole length $L_t$. The correlation
length $\xi_t(L_s)=1/m(L_s)$, defined for an infinitely long cylindrical
volume becomes much larger for $\lambda_0 \to 0$ than any finite $L_t$ .  
For the ``truly $\delta$-regime'' calculation one should study 
the dynamics of the spatially constant slow modes, which is described 
by the quantum mechanical 
rotator \cite{Leutwyler:1987ak,Hasenfratz:1993vf,Hasenfratz:2009mp}.  
Of course, the two approaches should lead to the same result for 
the mass gap, but one meets different technical difficulties in these 
two approaches.

\section{Procedure and checks}
It is straightforward to derive the Feynman rules and to write down the
corresponding Feynman diagrams. After that we separate the different $t^n$
contributions analytically from each graph.  This step is practically the same
for $d=2$ and $d>2$.  Evaluating the diagrams numerically provides a good
check for separating the different powers of $t$.  There are also other checks
for the final results:
\begin{itemize}
\item The consistency relations mentioned in section \ref{sct:ct} are
  satisfied.
\item The NLO result is known for general dimensions \cite{Hasenfratz:1993vf}.
\item The NNLO result for $d=2$ can be compared to the results of L\"uscher
  and Weisz \cite{LW_unpub,Shin:1996gi}
\item We also considered the same problem with Dirichlet-free boundary
  conditions. (This should obviously give the same result for the mass, but
  the corresponding transition amplitudes are different.)  We checked the
  result numerically for $d=2$, including the 4-derivative contributions.
\item Since the 4-derivative contributions were not calculated previously on
  the lattice for the 2d case, we compared the corresponding contributions to
  direct Monte-Carlo simulations for $d=2$, $L_s=3$ at sufficiently small
  $\lambda_0$ values.
\item We solved the problem for $n=2$ parametrizing the fields as
  $\mathbf{S}(x)=(\cos\phi(x),\sin\phi(x))$, including the 4-derivative
  contributions.
\end{itemize}

\section{The mass gap at finite lattice spacing}

Restoring the lattice spacing $a$ the mass gap reads
\begin{equation} \label{massgap} 
  m(L_s) = \frac{n_1\lambda_0^2 a^{d-2}}{2
    L_s^{d-1}} \left[ 1 + \lambda_0^2 c_2(a/L_s) + \lambda_0^4 c_3(a/L_s)+
    \lambda_0^4 \sum_{i=2}^{5} g_4^{(i)} d_3^{(i)}(a/L_s) \right] +
  \mathcal{O}(\lambda_0^8)
\end{equation}

The $n$-dependence of the coefficients is given by
\begin{align}
  c_2 & = c_{21} + c_{22} n_1 \,, \label{eq:c2}\\
  c_3 & = c_{31} + c_{32} n_1 + c_{33} n_1^2 \,, \label{eq:c3}\\
  d_3^{(i)} & = d_{31}^{(i)} + d_{32}^{(i)} n_1,\qquad
  i=2,3,4,5 \label{eq:d3}\,.
\end{align}
where we introduced the abbreviation
\begin{equation}
  n_1\equiv n-1\,.
\end{equation}
The above coefficients are expressed as lattice sums over spatial momenta and
are given in appendix \ref{app:analyticresults}.

\section{Results for $d=4$}
The expressions occurring in the previous section depend on the ratio $a/L_s$
only. In four space-time dimensions one has
\begin{align}
  &c_{2k}(a/L_s)=c_{2k0}+c_{2k1}\frac{a^2}{L_s^2} +
  {\mathcal{O}}\left(\frac{a^4}{L_s^4}\right)\,, \quad k=1,2\,,
  \label{c2ki}\\
  &c_{3k}(a/L_s)=c_{3k0}+c_{3k1}\frac{a^2}{L^2_s}+
  c_{3k2}\frac{a^4}{L_s^4}+c_{3k3}\frac{a^4}{L_s^4}\log\frac{L_s}{a} +
  {\mathcal{O}}\left(\frac{a^6}{L_s^6}\right)\,,
  \quad k=1,2,3\,,\label{c3ki} \\
  &d^{(i)}_{3k}(a/L_s)=D^{(i)}_{3k}+E^{(i)}_{3k}\frac{a^4}{L_s^4} +
  {\mathcal{O}}\left(\frac{a^6}{L_s^6}\right)\,, \quad i=2,3,4,5;\quad
  k=1,2\,.\label{d3ki}
\end{align}
In eq.~\eqref{massgap} these coefficients are multiplied by
$\lambda_0^2=1/F_0^2 a^2$ and $\lambda_0^4=1/F_0^4 a^4$, respectively.
Therefore the omitted terms correspond to lattice artifacts. 
The resulting terms with power-like and logarithmic singularities in $a$ 
will be absorbed by the renormalization of the lattice bare parameters. 
The numerical values of the coefficients are listed in
Tables~\ref{c21table}-\ref{d31table} in appendix \ref{app:numericalvalues}.

\subsection{Discussion on the numerical values of appendix
  \ref{app:numericalvalues}}

We have calculated the lattice sums in appendix \ref{app:analyticresults} 
to a quadruple precision for different lattice sizes and fitted 
the $a/L_s$-dependence.  
In order to find out a reliable set of terms one can vary
the minimum value of $L_s$ and include fewer or more higher powers like
$a^6/L_s^6$, $a^8/L_s^8$, etc. The coefficients should be stable under such
variations, and we can estimate the precision of the values, which was 10
digits or better.  Note, that we know the coefficients of the logarithmic
terms from DR, hence fixing these parameters in the fit allows to determine the
remaining ones to better accuracy.

The coefficient $c_{211}=-c_{221}$ is the shape coefficient $\beta_1^{(3)}$
for a three dimensional cubic box, given in \cite{Hasenfratz:1989pk}.  
The NLO finite size effect was first calculated in \cite{Hasenfratz:1993vf}.

The logarithms originate from the double sums in \eqref{c31a} and
\eqref{c32a}.  To obtain the leading $L_s$-dependence we expand them 
for small $\mathbf{k_1}$, $\mathbf{k_2}$,
\begin{align}
  \label{c31e}
  &c_{31} = \frac{1}{2V_s^2}\sum_{\mathbf{k_1 k_2}}{}^{''}
  \frac{s_1^2}{(s_1+s_2+s_3)s_1 s_2 s_3} + \ldots\\
  &c_{32} = -\frac{1}{2V_s^2}\sum_{\mathbf{k_1 k_2}}{}^{''}
  \frac{s_1^2}{(s_1+s_2+s_3)s_1s_2s_3} + \ldots\,,
\end{align}
where we expanded in $s_i \equiv \sinh\mu_i$. (See appendix
\ref{app:propagator} for notations.)  The leading singularities of the double
sums are related to those in the sum
\begin{multline}
  \label{Delta}
  \Delta \equiv \sum_x \Box_0 G(x) G(x)^2 = \\
  = \frac{1}{4V_s^2} \sum_{\mathbf{k_1 k_2}}{}^{''}
  \frac{(1+\mathrm{e}^{-\mu_1-\mu_2-\mu_3})(\cosh\mu_1-1)}{%
    (1-\mathrm{e}^{-\mu_1-\mu_2-\mu_3}) \sinh\mu_1 \sinh\mu_2 \sinh\mu_3}
  -\frac14 \Sm_1^2 - \frac{1}{8 V_s} \Sm_2
  \\
  = \frac{1}{4V_s^2} \sum_{\mathbf{k_1 k_2}}{}^{''}
  \frac{s_1^2}{(s_1+s_2+s_3)s_1 s_2 s_3} + \ldots\,.
\end{multline}
Therefore, one has
\begin{align}
  & c_{31} = 2 \Delta + \ldots\,,\\
  &c_{32} = -2 \Delta + \ldots\,.
\end{align}
According to the calculation in DR \cite{Hasenfratz:2009mp} the logarithmic
part of $\Delta$ is given by
\begin{equation}
  \int\! dx \, \Box_0 G(x) G(x)^2 = 
  -\frac{10}{3} \frac{q}{16\pi^2} \frac{\log L_s}{L_s^4} + \ldots \,,
\end{equation}
where $q=0.837536910696$ (cf. \eqref{eq:q}).  The coefficient of the
logarithmic term in $c_{31}$ is, therefore, $-0.0353584296400$ in agreement
with the direct fits given in Table~\ref{c31table} in appendix
\ref{app:numericalvalues}.

\subsection{Renormalization for $d=4$}
The additive renormalization of the 4-derivative operators in
eq.~\eqref{eq:A4total} are removed by setting
\begin{equation}
  c^{(i)}=\frac{2}{F_0^2a^4}
  \left( D_{31}^{(i)}+n_1 D_{32}^{(i)}\right) \,,\quad 
  i=2,3,4,5.
\end{equation}
The singular part of the $a$-dependence of the mass gap can be removed by the
renormalization of the bare lattice parameters as
\begin{align}
  &\frac{1}{F_0^2}= \frac{1}{F^2} \left[ 1 
    +\frac{{b}_1}{F^2a^2}+\frac{{b}_2}{F^4a^4}+
  \mathcal{O}\left(\frac{1}{F_0^6a^6}\right) \right] \label{eq:F0ren} \\
  &g_4^{(i)}={g}_{40}^{(i)}+{g}_{41}^{(i)}\log aM={g}_{41}^{(i)}\log
  aM_i,\quad i=2,3\,, \label{eq:g4ren}
\end{align}
where $M$ is a scale\footnote{In order to avoid confusion with DR we do not
  use $\Lambda$ or $\mu$.}.  Like in DR we introduce the individual scales
$M_2$ and $M_3$. Note that $g_4^{(4)}$ and $g_4^{(5)}$ do not need
renormalization to this order.  
The coefficients in eq.~\eqref{eq:F0ren} are expressed through
the infinite volume limits of coefficients in eqs.~\eqref{eq:c2},
\eqref{eq:c3} as
\begin{align}
  b_1 &= -c_2(0) = G(1,\mathbf{0})-n_1 G(0) = -\frac{1}{2d}-(n-2) G(0) \,,
  \label{b1} \\
  b_2 &= -c_3(0) + 2 c_2(0)^2 \,. \label{b2}
\end{align}
Their  numerical values for $d=4$ are 
\begin{align}\label{eq:b_1etc}
  b_1 &=0.029933390231-0.154933390231 \, n_1\,,\\
  b_2 &=0.001425585601-0.019893440100\, n_1+0.024004355409 \, n_1^2\,.
\end{align}
Note that there are several nontrivial relations that should be satisfied to
be able to absorb the singular cutoff dependence into the renormalized
couplings -- these are satisfied numerically to the expected accuracy,
providing an additional check.

The renormalization of $g_4^{(2)}$ and $g_4^{(3)}$ agrees with the result 
of ChPT in dimensional regularization \cite{Gasser:1983yg}, taking into 
account eq.~\eqref{g23-l12}\footnote{In ref.~\cite{Gasser:1983yg} 
  only the $n=4$ result is given but it is straightforward to restore 
  the general $n$ dependence.}.  
On the lattice (with Euclidean action) we have
\begin{align}
  g^{(2)}_{41}&=\frac{1}{16\pi^2}\left(\frac{14}{3}-2n_1\right)\,,\\
  g^{(3)}_{41}&=-\frac{1}{16\pi^2}\frac{8}{3}\,.
\end{align}

Note that, in contrast to the DR, the pion decay constant $F_0$ renormalizes
on the lattice. The NLO coefficient $b_1$ in \eqref{eq:F0ren} has been
calculated earlier in \cite{Hasenfratz:1993vf}.

ChPT in the $p$-regime with lattice regularization was studied earlier
by Shushpanov and Smilga \cite{Shushpanov:1998ms} for the 4d O(4) case,
who obtain, besides other 1- and 2-loop results,
also the renormalization of $F_0$ to NLO.
However, their result disagrees with ours --  they obtain
(for $n=4$) $b_1=-2 G(0)$, i.e. the $-1/(2d)$ term of eq.~\eqref{b1}
is missing. It is easy to verify its presence for the O(2) case 
in the same way as done in \cite{Shushpanov:1998ms} 
(from the current-current correlator)
when one uses the parametrization by the angular variable.
In this case the action is given by 
$\lambda_0^{-2} \sum_{x,\mu} (1-\cos(\lambda_0\partial_\mu \phi(x)))$
while the conserved current is
$A_\mu(x) =\lambda_0^{-2} \sin(\lambda_0\partial_\mu \phi(x))$.
In this case one gets $b_1=-1/(2d)$, as expected.

\subsection{The moment of inertia}

The isospin dependence of the lowest excitations up to (and including) NNLO
corrections are given by the rotator spectrum
\cite{Floratos:1984bz,Hasenfratz:2009mp}, i.e. by
\begin{equation}
  E_l=\frac{l(l+2)}{2\Theta},\quad l=0,1,2,\ldots\,.
\end{equation}
where $\Theta$ is the moment of inertia.  We have also calculated the
$E_2-E_0$ gap with the method presented here, and found agreement with this
expectation.  The moment of inertia is given by
\begin{equation}
  \begin{split} \label{thetafinal}
    \Theta=L_s^{3}F^2{\bigg[}1
    &+\frac{1}{F^2L_s^2}0.225784959441 \, (n-2)\\
    &+\frac{1}{F^4L_s^4}(-0.0692984943 +0.0101978424 \, n)\\
    &-\frac{1}{F^4L_s^4}0.007071685925\left[\left( 3n-10 \right)\log M_2L_s+ 
      2n\log M_3L_s\right]\\
    &-\frac{g_{4}^{(4)}}{F^4L_s^4}[-0.55835794046\, (n+1)]\\
    &-\frac{g_{4}^{(5)}}{F^4L_s^4}[0.55771822866 - 1.11639602502\,n]
    +\mathcal{O}\left(\frac{1}{F^6L_s^6}\right){\bigg{]}}.
  \end{split}
\end{equation}
The scales $M_2$ and $M_3$ are related to the corresponding scales
$\Lambda_1$, $\Lambda_2$ in dimensional regularization
(cf. eq.~\eqref{g23-l12}). However, to get this relation, and the values of the
coefficients $g_{4}^{(4)}$ and $g_{4}^{(5)}$ (needed to restore Lorentz
symmetry) one needs to relate the lattice regularization to the dimensional
one.  This step remains still to be done.  The resulting uncertainty is a term
$\mathrm{const}/(F^4 L_s^4)$.  The remaining terms are in agreement with the
result of ref.~\cite{Hasenfratz:2009mp} where the calculation was performed in
DR for the $n=4$ case.

\section{Results for $d=3$}
In three space-time dimensions the coefficients of \eqref{eq:c2}-\eqref{eq:d3}
depend on $a/L_s$ as
\begin{align}
  &c_{2k}(a/L_s)=c_{2k0}+c_{2k1}\frac{a}{L_s}+\ldots,\quad k=1,2\,,
  \label{c2ki3d}\\
  &c_{3k}(a/L_s)=c_{3k0}+c_{3k1}\frac{a}{L_s}+c_{3k2}\frac{a^2}{L_s^2}+\ldots,
  \quad k=1,2,3\,,\\
  &d^{(i)}_{3k}(a/L_s)=D^{(i)}_{3k}+\mathcal{O}\left(\frac{a^3}{L_s^3}\right),
  \quad i=2,3,4,5;\quad k=1,2\,.\label{d3ki3d}
\end{align}
The corresponding values are given in Tables~\ref{c21table3d}-\ref{d31table3d}
in appendix \ref{app:numericalvalues}.

\subsection{Renormalization and the moment of inertia for $d=3$}
The 4-derivative interactions only contribute to order $1/L_s^3$.  (Note that
the corresponding couplings $g_4^{(i)}$ are dimensionful, in contrast to
$d=4$).  They are also not needed for the renormalization here.  Because the
theory is non-renormalizable, it is expected that they are necessary to absorb
divergences at higher orders.  The spin stiffness renormalizes as
\begin{equation}
  \frac{1}{\rho_0}=\frac{1}{\rho}\left(1+\frac{{b}_1}{\rho a}+
    \frac{{b}_2}{\rho^2 a^2}+\ldots\right)\,.
\end{equation}
The corresponding coefficients are still given by eqs.~\eqref{b1},\eqref{b2}
while their numerical values are
\begin{align}
  &b_1=    0.0860643431920- 0.252731009859 \, n_1\,, \\
  &b_2=0.0102138509611 - 0.0659002864141\, n_1 + 0.0638729633447\, n_1^2
\end{align}
and the moment of inertia is
\begin{equation}
  \Theta=\rho L_s^2\left[
    1+\frac{n-2}{\rho L_s} 0.310373220693
    -\frac{n-2}{\rho^2 L_s^2}0.000430499941
    +\mathcal{O}\left(\frac{1}{L_s^3}\right)  \right].
\end{equation}      

The coefficient of the NNLO term has been estimated a long time ago from
finite temperature simulations in the spin $\frac12$ quantum Heisenberg model
\cite{Bea96} and the result cited there is not consistent with our
calculations.  The discrepancy of parameters obtained from those
finite-temperature data with other high precision measurements (about two per
mille, but statistically significant) has been observed already in
\cite{Ger09} and it was attributed to unaccounted finite-temperature
corrections.  A new measurement \cite{FJJ} of the staggered susceptibility at
much lower temperatures agrees with our very small coefficient of the $\propto
1/L_s^2$ term.

\section{Summary}

We calculated the mass gap in the O($n$) effective field theory in a cubic
spatial box of size $L_s$ for 3 and 4 space-time dimensions to NNLO, using
lattice regularization.  The renormalization of the bare lattice couplings is
performed, however, the connection of the 4-derivative couplings
with the $\overline{\mathrm{MS}}$ scheme is not established yet. 
This affects only the $\mathrm{const}/F^4 L_s^4$ term in the NNLO term 
in $d=4$ (relevant to QCD with two massless quarks), not the logarithmic terms, 
where we reproduced the result of \cite{Hasenfratz:2009mp}. 
Note that the effect of including a small explicit
symmetry breaking (a small quark mass) has been done to LO in
\cite{Leutwyler:1987ak}, and to NLO recently in \cite{Weingart}.  In $d=3$
(relevant for 2+1 dimensional spin $\frac12$ quantum Heisenberg model) our
calculation is complete to NNL order since the 4-derivative operators
do not contribute to this order.

{\bf Acknowledgments} The authors are indebted for useful discussions with
J. Balog, G. Colangelo, P. Hasenfratz, F.-J. Jiang, J. Kuti, H. Leutwyler,
M. Weingart, P. Weisz, and U.-J. Wiese.  This work is supported in part by the
Schweizerischer Nationalfonds.  The authors acknowledge support by DFG project
SFB/TR-55.  The ``Albert Einstein Center for Fundamental Physics'' at Bern
University is supported by the ``Innovations-und Kooperationsprojekt C-13'' of
the Schweizerischer Nationalfonds.

\begin{appendix}
  \section{The propagator}\label{app:propagator}

  We expand the leading action in the $\vec{\pi}$-fields and write it as
  \begin{equation}
    A_2=\frac{F^2}{2}\sum_x \partial_\mu\mathbf{S}_x \cdot 
    \partial_\mu\mathbf{S}_x = \frac{1}{2}\sum_{x,y} \rho(x,y) \vec{\pi}_x 
    \vec{\pi}_{y}+{\mathcal{O}}\left(\pi^4\right).
  \end{equation}
  The kernel $\rho(x,y)$ decomposes into the spatial and time direction
  \begin{equation}
    \rho(x,y)=\sum_{\mu=1}^{d-1} \rho_s(x_\mu-y_\mu) + \rho_0(x_0,y_0).
  \end{equation}
  The spatial kernel $\rho_s(x)$ is up to the sign the standard
  one-dimensional lattice Laplace operator
  \begin{equation}
    \rho_s(x)= 2\delta_{x,0} -\delta_{x,1}-\delta_{x,L_s-1}, 
    \quad x=0,1,\dots ,L_s-1,
  \end{equation}
  and periodically continued for $x\ge L_s$ and $x<0$.  It is convenient to
  write $\rho_0(x_0,y_0)$ for free boundary conditions in matrix form
  \begin{equation} \label{eq:rho0} \rho_0 =
    \begin{pmatrix}
      1  & -1 & 0  & \ldots & 0  &  0 & 0  \\
      -1 & 2  & -1 & \ldots & 0  &  0 & 0  \\
      \hdotsfor{7}                         \\
      0  &  0 & 0  & \ldots & -1 &  2 & -1 \\
      0 & 0 & 0 & \ldots & 0 & -1 & 1
    \end{pmatrix}.
  \end{equation}
  Consider the partial Fourier transform of the kernel
  \begin{equation}
    \tilde{\rho}(x_0,y_0,\mathbf{k})=\sum_\mathbf{x}
    \rho(x_0,y_0,\mathbf{x}-\mathbf{y})
    \mathrm{e}^{i\mathbf{k}(\mathbf{x}-\mathbf{y})}= 
    \rho_0(x_0,y_0) + \delta_{x_0, y_0}\tilde{\rho}_s(\mathbf{k}) \,,
  \end{equation}
  where
  \begin{equation}\label{mudef}
    \tilde{\rho}_s(\mathbf{k}) =
    \sum_{i=1}^{d-1} 4 \sin^2(k_i/2) \equiv 4 \sinh^2(\mu(\mathbf{k})/2).
  \end{equation}
  This equation defines $\mu(\mathbf{k})$ which is denoted by $\mu$ in the
  following. It will be useful to separate the $\mathbf{k}=\mathbf{0}$
  contribution in the propagator
  \begin{equation} \label{G0G1}
    \begin{split}
      G(x,y) \equiv G(x_0,y_0, \mathbf{x}-\mathbf{y}) &=
      \frac{1}{V_s}\sum_{\mathbf{k}}G(x_0,y_0; \mu(\mathbf{k}))
      \mathrm{e}^{i \mathbf{k}(\mathbf{x}-\mathbf{y})}\\
      &= \frac{1}{V_s} g_0(x_0,y_0) + G_1(x,y) =G_0(x_0,y_0)+G_1(x,y).
    \end{split}
  \end{equation}
  The 1-dimensional massless propagator is the inverse of $\rho_0$ in
  \eqref{eq:rho0} with the zero mode left out (the boundaries are at $t=\pm
  T$)
  \begin{equation}
    \label{G0freefree}
    g_0(x_0,y_0)=-\frac{1}{2}|x_0-y_0|+\frac{1}{2(2T+1)}(x_0^2+y_0^2)
    +\frac{T(T+1)}{3(2T+1)}.
  \end{equation}
  and
  \begin{equation}\label{eq:G1}
    G_1(x,y) =
    \frac{1}{V_s}\sum_{\mathbf{k}}{}^{'} G(x_0,y_0; \mu)
    \mathrm{e}^{i \mathbf{k}(\mathbf{x}-\mathbf{y})}.
  \end{equation}
  The primed sum symbol means that the $\mathbf{k}=\mathbf{0}$ zero mode is
  excluded. The function $ G(t,t';\mu)$ is the massive propagator for the 1d
  case with free b.c.:
  \begin{multline} \label{eq:GFF1} G(t,t';\mu) =
    \frac{1}{2\sinh\mu \, (1-\mathrm{e}^{-2\mu(2T+1)})} \\
    \left\{ \mathrm{e}^{-\mu|t-t'|} + \mathrm{e}^{-\mu(4T+2-|t-t'|)} +
      \mathrm{e}^{-\mu(2T+1+t+t')} + \mathrm{e}^{-\mu(2T+1-t-t')} \right\}.
  \end{multline}
  This function is exponentially small for $|t-t'|\gg L_s$.  In the limit
  $T\to\infty$ and when both time arguments are finite the propagator is
  simplified,
  \begin{equation} \label{eq:GFF2} G(t,t';\mu) \approx
    \frac{\mathrm{e}^{-\mu|t-t'|}}{2\sinh\mu }.
  \end{equation}

  The $\propto t^n$ dependence of a Feynman graph for $C(t)$ comes from the
  time-like part $G_0(x_0,y_0)$ in \eqref{G0G1}.

  \section{Analytic results}\label{app:analyticresults}
  \subsection{Contributions from the leading action}
  We introduce the notations
  \begin{align}
    \Sm_n & = \frac{1}{V_s} \sum_\mathbf{k}{}^{'}\frac{1}{\sinh^n \mu} \,, \\
    \Em_n & = \frac{1}{V_s} \sum_\mathbf{k}{}^{'}
    \frac{1 - \mathrm{e}^{-\mu}}{\sinh^n \mu} \,, \\
    \Cm_n & = \frac{1}{V_s} \sum_\mathbf{k}{}^{'} \frac{\cosh\mu - 1}{\sinh^n
      \mu} \,.
  \end{align}
  The coefficients read:
  \begin{equation}
    \label{c21}
    c_{21}(a/L_s) 
    = \frac{1}{2V_s} - \SUMV \frac{\ei}{2 \si} \,,
  \end{equation}

  \begin{equation}
    \label{c22}
    c_{22}(a/L_s) 
    = \SUMV\frac{1}{2 \si} \,,
  \end{equation}

  \begin{multline}
    \label{c31a}
    c_{31}(a/L_s) = \frac{1}{2V_s^2} - \frac{1}{12} \Sm_1 + \frac{1}{12 V_s} (
    -10 \Sm_1 - 3 \Em_3 + 6 \Em_2
    + 12 \Em_1 ) \\
    +\frac14 \Em_1 ( \Cm_3 + 2 \Em_1 - 2 \Sm_1 )
    - \frac{1}{4(d-1)}  \Cm_3 \Cm_1 \\
    + \frac{1}{48 V_s^2}\sum_{\mathbf{k_1 k_2}}{}^{''}
    \frac{1}{(1-\e^{-\mu_1-\mu_2-\mu_3}) \sinh\mu_1\sinh\mu_2\sinh\mu_3}
    \left[ \e^{\mu_1+\mu_2} \right. \\
    \left.  +10 \e^{\mu_1-\mu_2}-13\e^{-\mu_1}+14\e^{-2\mu_1-\mu_2}
      -11\e^{-\mu_1-\mu_2}-\e^{-2\mu_1-2\mu_2-\mu_3}\right]\,,
  \end{multline}
  \begin{multline}
    \label{c32a}
    c_{32}(a/L_s) = \frac14 ( \Sm_2 - \Em_3 )
    + \frac{1}{24 V_s} ( 15 \Sm_2 + 2 \Sm_1 + 10 \Em_1 )\\
    + \frac{1}{24}(\Em_1 - \Sm_1)( 11\Sm_1 + \Em_1) -\frac16 \Sm_1^2 -
    \frac{1}{12} \Sm_1 \Cm_1 -\frac14 \Cm_1 \Cm_3
    \\
    +\frac{1}{24(d-1)} (\Cm_1 + 6 \Sm_1 - 6 \Cm_3)\Cm_1
    \\
    +\frac{1}{96V_s^2}\sum_{\mathbf{k_1 k_2}}{}^{''}
    \frac{1}{(1-\e^{-\mu_1-\mu_2-\mu_3}) \sinh\mu_1\sinh\mu_2\sinh\mu_3}
    \left[ \e^{2\mu_1}\right. \\
    +12-13\e^{-2\mu_1}-9\e^{-\mu_1-\mu_2+\mu_3}
    +8\e^{-\mu_1-\mu_2-\mu_3}+\e^{-\mu_1-\mu_2-3\mu_3} \\
    \left.-4\e^{-2\mu_1-2\mu_2}+4\e^{-2\mu_1-2\mu_2-2\mu_3}\right]\,,
  \end{multline}
  (The double prime in the sum over $\mathbf{k_1}$, $\mathbf{k_2}$ means the
  condition that $\mathbf{k_1}\ne \mathbf{0}$, $\mathbf{k_2}\ne \mathbf{0}$,
  and $ \mathbf{k_3} = \mathbf{k_1}-\mathbf{k_2}\ne \mathbf{0}$.)
 
  \begin{equation}
    \label{c33}
    c_{33}(a/L_s) = -\frac{1}{24 V_s^2} + \left( \SUMV \frac{1}{2\sinh\mu} 
    \right)^2. 
  \end{equation}

\subsection{Contributions from the next-to-leading action}
In this section we give the contributions of the 4-derivative interactions,
without the contribution of the subtracted terms proportional to $c^{(i)}$ in
equation \eqref{eq:A4total}. The effect of these terms is discussed in the
renormalization chapter.  The contributions of the interactions 2,3,4 depend
only on the expression
\begin{equation}
  \Omega_0\doteq\partial_0^u\partial_0^vG_{uv}{}_{\big|_{v=u}}= 
  \frac{1}{V_s}+\SUMV \frac{1-\ei}{\sinh\mu}=\frac{1}{V_s}+\Em_1.
\end{equation}
They are given by \vspace{-0.4cm}
\begin{center}
  \begin{minipage}[h]{0.45\linewidth}
    \begin{align}
      d_{31}^{(2)} &= -2 \Omega_0\\
      d_{31}^{(3)} &= -1 - \Omega_0\\
      d_{31}^{(4)} &= -\frac{2d}{d+2}\left(\Omega_0-\frac{1}{d}\right)
    \end{align}
  \end{minipage}
  \begin{minipage}[h]{0.45\linewidth}
    \begin{align}
      \label{eq:d2_22}
      d_{32}^{(2)} &= -1 \\
      \label{eq:d3_22}
      d_{32}^{(3)}&=  -\Omega_0 \\
      \label{eq:d4_22}
      d_{32}^{(4)} &= -\frac{d}{d+2}\left(\Omega_0-\frac{1}{d}\right) \,.
    \end{align}
  \end{minipage}
\end{center}
The contribution of the fifth operator is more complicated. Using the labels
introduced in section \ref{sct:Action} we write
\begin{equation}
  d_{3i}^{(5)}= d_{3i}^{(5a)}-\frac{1}{d+2}\left[ 2d_{3i}^{(5b)} 
    + d_{3i}^{(5c)}\right],\quad i=1,2.
\end{equation}
We obtain
\begin{equation}
  \label{m51A}
  d_{31}^{(5a)}= -\frac{1}{V_s} - \SUMV \frac{(4+3\ei+\eii)\eii}{(1+\ei)^3} +
  \SUMV \frac{1}{2\siii} \Cisq\,,
\end{equation}
\begin{equation}
  \label{m52A}
  d_{32}^{(5a)}= -\frac{1}{2V_s} - \SUMV \frac{(3+\ei)\eii}{(1+\ei)^3}
  - \SUMV \frac{\ci}{2\siii} \Cisq\,,
\end{equation}

\begin{equation}
  \label{mI1A}
  d_{31}^{(5b)}= -\frac{1}{V_s} - \SUMV \ei\,,
\end{equation}

\begin{equation}
  \label{mI2A}
  d_{32}^{(5b)} = -\frac12\,,
\end{equation}

\begin{multline}
  \label{mIII1A}
  d_{31}^{(5c)} = -\frac{1}{V_s} + \frac{1}{V_s} \SUM
  \frac{(3-8\ei-4\eii+\eiiii)\ei}{(1+\ei)^3}
  \\
  - \frac{1}{V_s} \SUM \frac{(2-\ei-2\eii-\eiii)\ei}{(1+\ei)^2\si} \Cisq
  \\
  + \frac{1}{V_s} \SUM \frac{1}{2\siii} \left[\CS - (\ci-1)^2\right] \,,
\end{multline}

\begin{multline}
  \label{mIII2A}
  d_{32}^{(5c)}= -\frac{1}{2V_s} - \SUMV
  \frac{1-3\ei+13\eii-\eiii-2\eiiii}{2(1+\ei)^3}
  \\
  - \SUMV \frac{(1-2\ei-\eii)\ei}{(1+\ei)^2\si} \Cisq
  \\
  - \frac{1}{V_s} \SUM \frac{\ci}{2\siii} \left[ \CS - (\ci-1)^2\right].
\end{multline}

\section{Numerical values}\label{app:numericalvalues}

\subsection{Numerical values for $d=4$}
The values for the coefficients occurring in equation \eqref{c2ki} and the
following are collected in Tables~\ref{c21table}-\ref{d31table} where we
introduced the notations

\begin{equation} \label{eq:q}
  \begin{split}
    q = 0.837536910696\,, \qquad
    & r = 0.9764866840\,, \\
    v_0 = -0.03163189123\,, \qquad & v_1 = 0.01546809528 \,.
  \end{split}
\end{equation}

\begin{center}
  \renewcommand{\baselinestretch}{1.2}
  \begin{table}[H]
    \begin{center}
      \begin{tabular}{|l|l|l|}
        \hline
        $k$ & $c_{21k}$  & $c_{22k}$ \\
        \hline
        $0$ &-0.029933390231  & 0.154933390231 \\
        \hline
        $1$ & 0.225784959441  & -0.225784959441 \\
        \hline 
      \end{tabular}\caption{Coefficients at NLO, $d=4$. \label{c21table}}
    \end{center}
  \end{table}
\end{center}
\begin{center}
  \renewcommand{\baselinestretch}{1.2}
  \begin{table}[H]
    \begin{center}
      \begin{tabular}{|l|l|l|l|}
        \hline
        $k$ & $c_{31k}$  & $c_{32k}$ & $c_{33k}$ \\
        \hline
        $0$ & 0.000366430100  & 0.001342713582 &  0.024004355408   \\
        \hline
        $1$ & -0.013517018599  & 0.083480277057 & -0.069963258459    \\
        \hline
        $2$ &0.0961483532   & -0.051187951 & 0.108634522    \\
        \hline
        $3$ &-0.035358429639  & 0.035358429639. & 0   \\
        \hline 
      \end{tabular}\caption{Coefficients at NNLO, $d=4$. \label{c31table}}
    \end{center}
  \end{table}
\end{center}

\begin{center}
  \renewcommand{\baselinestretch}{1.2}
  \begin{table}[H]
    \begin{center}
      \begin{tabular}{|c|c|c||c|c||c|c||c|c|}
        \hline
        $k$ & $D^{(2)}_{3k}$ & $E^{(2)}_{3k}$ & $D^{(3)}_{3k}$& $E^{(3)}_{3k}$
        & $D^{(4)}_{3k}$  & $E^{(4)}_{3k}$  
        & $D^{(5)}_{3k}$& $E^{(5)}_{3k}$  \\
        \hline
        $1$& $-\frac{1}{2}$  & $-2q$ & $-\frac{5}{4}$& $-q$  & 0 & 
        $-\frac{4q}{3}$  & $v_0$& $-\frac{5}{4}q+\frac{1}{2}r$         \\
        \hline
        $2$ & $-1$  & $0$ & $ -\frac{1}{4}$& $-q$& 0 & $ -\frac{2q}{3}$ & 
        $v_1$& $-\frac{3}{4}q-\frac{1}{2}r$    \\
        \hline
      \end{tabular}\caption{Coefficients for the 4-derivative contributions 
        at NNLO, $d=4$. \label{d31table}}
    \end{center}
  \end{table}
\end{center}

\subsection{Numerical values for $d=3$}
The values for the coefficients occurring in equation \eqref{c2ki3d} and the
following are given in Tables~\ref{c21table3d}-\ref{d31table3d}.

\begin{center}
  \renewcommand{\baselinestretch}{1.2}
  \begin{table}[H]
    \begin{center}
      \begin{tabular}{|l|l|l|}
        \hline
        $k$ & $c_{21k}$  & $c_{22k}$ \\
        \hline
        $0$ &-0.086064343192  & 0.252731009859 \\
        \hline
        $1$ & 0.310373220693  & -0.310373220693 \\
        \hline 
      \end{tabular}\caption{Coefficients at NLO, $d=3$. 
        \label{c21table3d}}
    \end{center}
  \end{table}
\end{center}

\begin{center}
  \renewcommand{\baselinestretch}{1.2}
  \begin{table}[H]
    \begin{center}
      \begin{tabular}{|l|l|l|l|}
        \hline
        $k$ & $c_{31k}$  & $c_{32k}$ & $c_{33k}$ \\
        \hline
        $0$ &  0.004600291377& -0.021104227057 & 0.063872963344    \\
        \hline
        $1$ &  -0.053424134767&  0.210306009764& -0.156881874998    \\
        \hline
        $2$ & 0.095901036182 &-0.192232572305& 0.096331536123      \\
        \hline 
      \end{tabular}\caption{Coefficients at NNLO, $d=3$. 
        \label{c31table3d}}
    \end{center}
  \end{table}
\end{center}

\begin{center}
  \renewcommand{\baselinestretch}{1.2}
  \begin{table}[H]
    \begin{center}
      \begin{tabular}{|c|c||c||c||c|}
        \hline
        $k$ & $D^{(2)}_{3k}$ & $D^{(3)}_{3k}$ & $D^{(4)}_{3k}$ & $D^{(5)}_{3k}$ \\
        \hline
        $1$&   $-\frac{2}{3}$ & $-\frac{4}{3}$  &  0 &  $0.038725856392$ \\
        \hline
        $2$&  $-1$ &    $-\frac{1}{3}$  &      0   &    $0.008715670880$ \\
        \hline
      \end{tabular}\caption{Coefficients for the 4-derivative 
        contributions at NNLO, $d=3$. \label{d31table3d}}
    \end{center}
  \end{table}
\end{center}

\end{appendix}


\end{document}